\RequirePackage{fix-cm}
\documentclass[epjc3]{svjour3}  
\smartqed  
\RequirePackage{graphicx}

\usepackage[margin=2.0cm]{geometry}

\usepackage{amssymb}
\usepackage[numbers]{natbib}

\usepackage{latexsym}
\usepackage{revsymb}
\usepackage{multirow}
\usepackage{color}
\usepackage[usenames,dvipsnames,svgnames,table]{xcolor}

\usepackage{graphicx}
\usepackage{epsfig}  
\usepackage{epsf}    
\usepackage{dcolumn}
\usepackage{bm}
\usepackage{dcolumn}
\usepackage{textcomp}
\usepackage{float}
\usepackage{subfig}

\usepackage{lineno}
\modulolinenumbers[5]

\usepackage[hyperfootnotes=false]{hyperref}
\usepackage[all]{hypcap}
\usepackage{pdftricks}
\usepackage{amsmath}
\usepackage{bm}
\usepackage{amssymb}
\usepackage{amsfonts}

\begin{document}




\title{Studying the neutrino wave-packet effects at medium-baseline reactor neutrino oscillation experiments and the potential benefits of an extra detector}
\author{Zhaokan Cheng\thanksref{e1,addr1}
        \and
        Wei Wang\thanksref{e2,addr2, addr3}
        \and
        Chan Fai Wong\thanksref{e3,addr2} 
        \and
        Jingbo Zhang\thanksref{e4,addr1}
}

\thankstext{e1}{14b911018@hit.edu.cn}
\thankstext{e2}{wangw223@sysu.edu.cn}
\thankstext{e3}{wongchf@sysu.edu.cn}
\thankstext{e4}{jinux@hit.edu.cn}


\institute{School of Physics, Harbin Institute of Technology, Harbin, China \label{addr1}
           \and 
           School of Physics, Sun Yat-sen University, Guangzhou, China \label{addr2}
\and Sino-French Institute of Nuclear Engineering and Technology, Sun Yat-sen University, Zhuhai, China \label{addr3}
}

\date{}
          
\maketitle




\begin{abstract}


We examine the potential of the future medium-baseline reactor
neutrino oscillation (MBRO) experiments in studying neutrino
wave-packet impact. In our study, we treat neutrinos as wave
packets and use the corresponding neutrino flavor transition
probabilities. The delocalization, separation and spreading of the
wave packets lead to decoherence and dispersion effects, which modify
the plane-wave neutrino oscillation pattern, by amounts that depend on
the energy uncertainties in the initial neutrino wave packets. We find
that MBRO experiments
could be sensitive to the wave-packet impact, since the baseline is
long enough and also the capability of observing small corrections to
the neutrino oscillations due to excellent detector energy
resolution. Besides studying the constraints on the decoherence
parameter, we also examine the potential wave-packet impacts on the
precision of measuring $\theta_{12}$ and other oscillation parameters
in the future medium-baseline reactor neutrino oscillation
experiments. Moreover, we also probe the potential benefits of an
additional detector for studying such exotic neutrino physics.


\end{abstract}
\maketitle


\pagenumbering{arabic}

\section{Introduction}\label{sec1}
The plane-wave description of neutrino mixing and oscillation has been the standard picture for neutrino
oscillation\cite{Eliezer:1975ja} and all the parameters have been defined and analyzed based on such
a picture. Till now, this standard picture has been very consistent with the neutrino experimental data
\cite{Chan:2015mca, An:2016pvi}.
However, as neutrino production and detection are spatially localized, there must be finite intrinsic
energy/momentum uncertainties. A wave-packet description is
naturally expected to be more general and appropriate for a complete understanding of neutrino oscillations 
in reality \cite{ Chan:2015mca, An:2016pvi,
Kayser:1981ye,Kiers:1995zj,Giunti:2002xg,Dolgov:2005vj,Giunti:2007ry,Akhmedov:2009rb,Naumov:2010um}.
As neutrino physics is entering a precision stage, more
advanced detector technologies are becoming available, especially the
MBRO type experiments of resolving neutrino mass
hierarchy~(MH)\cite{Petcov, Zhan:2008id, Zhan:2009rs, Qian:2012xh,
Kettell:2013eos, Takaesu:2013wca, Li:2013zyd, Kim:2014rfa,
An:2015jdp}, which is made possible by an unexpectedly large value of
$\theta_{13}$ found by the current generation of short-baseline
reactor neutrino and long-baseline accelerator neutrino experiments
\cite{An:2012eh, Ahn:2012nd, Abe:2011fz, Abe:2013hdq, Adey:2018zwh} 
%
%
Thanks to their extradinary capability of measuring the multiple oscillation cycles, 
MBRO type detectors are expected to be also sensitive to potential damping
signatures resulted from various non-standard mechanisms of neutrino
flavour transitions, such as the neutrino wave-packet hypothesis
\cite{Chan:2015mca, An:2016pvi, Giunti:2007ry, Akhmedov:2009rb,
Beuthe:2001rc, Blennow:2005yk, Bernardini:2006ak,Naumov:2013vea,
Chatelain:2019nkf, Penacchioni:2019kix, Ciuffoli:2020flo}.
As the first MBRO project, the JUNO experiment consists of a large central detector with unprecedented
energy resolution (3\%/$\sqrt{E/\mathrm{MeV}}$), a water Cherenkov
detector and a muon tracker. The central detector is a liquid
scintillator (LS) detector within a target mass of 20 kton at a $\sim$
34.4 m fiducial volume. It is built in the Jinji town located at
$\sim$ 52.5 km from the Yangjiang Nuclear Power Plant (NPP) and
Taishan NPP, where 10 rector cores offer a combined thermal power of
$\sim$ 35.8 GW$_\text{th}$. With such a setup, the JUNO
experiment is believed to be a state-of-the-art platform to
identify the neutrino mass ordering and also perform the precision
measurements of various neutrino oscillation parameters, such as
$\theta_{12}$, $\Delta m^2_{21}$ and $\Delta m^2_{ee}$
\cite{An:2015jdp}. In the following, we will also discuss the
wave-packet impact on such precision measurements.

In this article, we apply a wave-packet treatment to neutrino
oscillations and probe the potential of MBRO experiments in studying the
neutrino wave-packet hypothesis. We examine the constraints on the
decoherence parameter ($\sigma_\mathrm{wp}$) at medium-baseline
reactor neutrino oscillation experiments and also investigate how does
the neutrino wavep-packet treatment affect the precision measurement
of oscillation parameters. 
This article is organized as follows. In section \ref{sec2}, we discuss the mechanism of neutrino
wave-packet treatment and briefly review the discussions in the
literature about this hypothesis. In section \ref{sec3}, we show the
resulting constraints on the neutrino wave-packet parameter from the
future MBRO experiment(s) and compare them with the current
constraints from Daya Bay. Then in section \ref{sec4}, we discuss the
wave-packet impacts on the precision measurement of oscillation
parameters. In section \ref{sec5}, we further discuss the potential of an extra
detector at MBRO experiment(s) on the studies of decoherence and
dispersion effects due to neutrino wave-packet treatment. At last, a
summary of our results and perspectives are presented in section \ref{sec6}.

\section{The neutrino wave-packet hypothesis}\label{sec2}
The plane-wave description of neutrino oscillation has been developed for almost 40 years \cite{Eliezer:1975ja}. However, as neutrino production and detection are spatially localized, there must be finite intrinsic energy/momentum uncertainties and a neutrino should be described by a wave packet. A wave-packet description is expected to be more general and appropriate for a complete understanding of neutrino oscillations 
in reality, each neutrino emitted from the source should have a mixture of energy / momentum states, and thus a wave-packet (WP) description is more self-consistent and appropriate, which leads to modification of the plane-wave neutrino oscillation probability by terms that depend on the energy / momentum width of the initial neutrino wave packet, $\sigma_\nu$ \cite{Kayser:1981ye,Kiers:1995zj,Giunti:2002xg,Dolgov:2005vj,Giunti:2007ry,Akhmedov:2009rb,Naumov:2010um, Chan:2015mca, An:2016pvi}. 
Massive neutrino should be described by a wave packet as it propagates freely \cite{Kiers:1995zj, Akhmedov:2009rb, Naumov:2010um}:

\begin{align}\label{wave_packet}
 |\nu_i(z,t) \rangle & = \int_{- \infty}^\infty \dfrac{dp}{\sqrt{2\pi}}\dfrac{1}{\sqrt{\sqrt{\pi}\sigma_\nu}}\mathrm{exp} \left[-\dfrac{(p-p_\nu)^2}{2\sigma_\nu^2}\right] \cdot \mathrm{exp}[i(pz-E_i(p)t)] |\nu_i \rangle, \\  
 |\nu_\alpha(z,t) \rangle & = \sum_{i} U_{\alpha i}^\ast |\nu_i(z,t) \rangle, \label{flavor_state}
\end{align}
where $|\nu_i \rangle$ is an energy eigenstate with energy $E_i$, $p_\nu$ is the mean momentum, $\sigma_\nu$ is the width of the wave packet in momentum space\footnote{Here, $\sigma_\nu$ is the effective uncertainty, with $1/\sigma_\nu^2 = 1/\sigma_\mathrm{prod}^2 + 1/\sigma_\mathrm{det}^2$, which has included both the production and detection neutrino energy uncertainties \cite{Giunti:2002xg, Giunti:2003ax, Beuthe:2001rc}. Moreover, we would like to point out that $\sigma_\mathrm{det}$ represents the energy uncertainty of detection at the microscopic level, i.e., that of the inverse-beta decay reaction. This is different from the detector energy resolution, which is determined by macroscopic parameters such as the performance of PMTs and geometry of the anti-neutrino detector, etc. In principle, the detector resolution is irrelevant for the size of the neutrino wave packets.}, assumed to be independent of the neutrino energy here, and $|\nu_\alpha \rangle$ is a neutrino flavor state. 

\subsection{The conventional decoherence effect due to separation of wave packets}\label{sec2.1}

In order to calculate the integral in Eq. (\ref{wave_packet}), the energy $E_i(p)$ has to be expanded around the mean momentum $p_\nu$. In most decoherence literatures, it is just expanded to first order as the higher order terms are expected to be strongly suppressed by the factors of $(\dfrac{m^2}{E_i^2})^n$:
\begin{equation}\label{E_expand_1st}
 E_i(p) \approx E_i(p_\nu) + v_i(p_\nu)(p-p_\nu),
\end{equation}
where $v_i(p_\nu) = \left. \dfrac{dE_i}{dp}\right|_{p = p_\nu} = p_\nu/E_i (p_\nu)$, which is the group velocity of the wave packet. Based on Eqs. (\ref{wave_packet}) and (\ref{E_expand_1st}), the neutrino oscillation probability of $\nu_\alpha \rightarrow \nu_\beta$ would be given by \cite{Fuji:2006mq}:
\begin{align}\label{Prob_inf_1st}
 P_{\nu_\alpha \rightarrow \nu_\beta}(L) & \approx \sum_{ij} \left[ U_{\alpha i}^\ast U_{\beta i}U_{\alpha j}U_{\beta j}^\ast \mathrm{exp}\left(-i\dfrac{2\pi L}{L^\mathrm{osc}_{ij}}\right)\right]  \mathrm{exp}\left(-\dfrac{L^2}{(L^\mathrm{coh}_{ij})^2} \right) \\
  \mbox{where } L^\mathrm{osc}_{ij} & \equiv \dfrac{4\pi E}{\Delta m^2_{ij}}, \quad \qquad L^\mathrm{coh}_{ij} \equiv \dfrac{L^\mathrm{osc}_{ij}}{\pi \sigma_\mathrm{wp}} = \dfrac{4E}{\Delta m^2_{ij} \sigma_\mathrm{wp}}, \quad \qquad \sigma_\mathrm{wp} \equiv \dfrac{\sigma_\nu}{E_i(p_\nu)} \approx \dfrac{\sigma_\nu}{E(p_\nu)}. \notag 
\end{align}
$L_\mathrm{coh}$ is the coherence length, which represents the distance where the decoherence effect becomes significant.

In the future medium-baseline reactor neutrino oscillation experiment(s), the $\bar\nu_e$ survival probability formula would be rewritten as
\begin{align}\label{Pee_1st}
 P_{\bar{e}\bar{e}} = & 1 - \frac{1}{2}\mathrm{cos}^4(\theta_{13})\mathrm{sin}^2(2\theta_{12})[1 - \mathrm{exp}(-\sigma_\mathrm{wp}^2\dfrac{(\Delta m^2_{21})^2 L^2}{16E^2})\mathrm{cos}(2\dfrac{\Delta m^2_{21}L}{4E})] \notag \\
          & - \frac{1}{2}\mathrm{sin}^2(2\theta_{13})\mathrm{cos}^2(\theta_{12})[1 - \mathrm{exp}(-\sigma_\mathrm{wp}^2\dfrac{(\Delta m^2_{31})^2 L^2}{16E^2})\mathrm{cos}(2\dfrac{\Delta m^2_{31}L}{4E})] \notag \\
          & - \frac{1}{2}\mathrm{sin}^2(2\theta_{13})\mathrm{sin}^2(\theta_{12})[1 - \mathrm{exp}(-\sigma_\mathrm{wp}^2\dfrac{(|\Delta m^2_{31}| - \Delta m^2_{21})^2 L^2}{16E^2})\mathrm{cos}(2\dfrac{(|\Delta m^2_{31}| - \Delta m^2_{21})L}{4E})].
\end{align}
Eqs. (\ref{Prob_inf_1st}) and (\ref{Pee_1st}) could be found in most decoherence literatures \cite{SigmaxRange, Fuji:2006mq, Blennow:2005yk}, which describe the decoherence effect due to the fact that different mass eigenstates travel with different speeds and they therefore gradually separate, reducing their interference and leading to a damping of neutrino oscillations. However, the quadratic correction to the neutrino energy and the decoherence effect due to delocalization (i.e., the spatial width of neutrino wave packet $\sigma_x$ being too large) \cite{Beuthe:2001rc, Chan:2015mca}  have not been taken into account yet. In the following subsection, we will derive the oscillation probability more precisely.

\subsection{The dispersion effect}\label{sec2.2}
In this subsection, we use the wave-packet treatment and approximations in Reference \cite{Chan:2015mca} to calculate the integral in Eq. (\ref{wave_packet}), which includes the second order correction to the neutrino energy.
\begin{equation}\label{E_expand}
 E_i(p) \approx E_i(p_\nu) + v_i(p_\nu)(p-p_\nu) + \dfrac{m_i^2}{2(E_i(p_\nu))^3}(p-p_\nu)^2,
\end{equation}
Conventionally, the last term in Eq. (\ref{E_expand}) is neglected since it is strongly suppressed by the factor $(\dfrac{m_i^2}{E_i^2})$. However, this term give rises to the dispersion of the wave packet and would alter the survival probability if $\sigma_\mathrm{wp}$ is large. The phenomenological consequence of this term is that the dispersion effect will partially compensate the 
decoherence effect due to the linear term in Eq. (\ref{E_expand}) and further modify the neutrino oscillation pattern. Then the neutrino flavor transition probabilities at baseline $L$ is given by: 

\begin{align}
 P_{\nu_\alpha \rightarrow \nu_\beta}(L) & \approx \sum_{ij} \left\{ U_{\alpha i}^\ast U_{\beta i}U_{\alpha j}U_{\beta j}^\ast \mathrm{exp}\left[-i\dfrac{2\pi L}{L^\mathrm{osc}_{ij}}\right] \right\} \notag \\
 & \left\{ \left(\dfrac{1}{1+y_{ij}^2}\right)^{\frac{1}{4}}\mathrm{exp}(-\lambda_{ij})\mathrm{exp}\left(\frac{-i}{2} \mbox{tan}^{-1} (y_{ij})\right) \mathrm{exp}(i\lambda_{ij}y_{ij}) \right\}, \label{Prob_inf1}  \\
                                         \mbox{where }& \lambda_{ij} \equiv \dfrac{x_{ij}^2}{1+y_{ij}^2},  \quad \qquad y_{ij} \equiv \dfrac{L}{L^\mathrm{dis}_{ij}},  \quad \qquad x_{ij} \equiv \dfrac{L}{L^\mathrm{coh}_{ij}}, \notag \\
                                        & L^\mathrm{osc}_{ij} \equiv \dfrac{4\pi E}{\Delta m^2_{ij}}, \qquad  L^\mathrm{coh}_{ij} \equiv \dfrac{4E}{\Delta m^2_{ij} \sigma_\mathrm{wp}}, \qquad  L^\mathrm{dis}_{ij} \equiv \dfrac{2E}{\Delta m^2_{ij} \sigma_\mathrm{wp}^2}, \notag \\
                                        & \sigma_\mathrm{wp} = \dfrac{\sigma_\nu}{E_i(p_\nu)} \approx \dfrac{\sigma_\nu}{E(p_\nu)} \notag.
\end{align}
The terms in the first bracket correspond to the standard plane-wave oscillation probabilities, and those in the second bracket represent the modifications due to wave-packet impact. The exp$(-\lambda_{ij})$ term corresponds to the decoherence effect due to the fact that different mass states propagate at different speeds $v_i(p_\nu)$ and they gradually separate and stop to interfere with each other, resulting in a damping of oscillations. The terms depending on $y_{ij}$ describe the dispersion effects and are dependent of the dispersion length(s) $L^\mathrm{dis}_{ij}$. Furthermore, 
$y_{ij}$ are proportional to $\sigma_\mathrm{wp}^2$, while $x_{ij} \propto \sigma_\mathrm{wp}$ only. Therefore, if $\sigma_\mathrm{wp} \ll 1$, the dispersion effect is expected to be more suppressed and negligible. Dispersion has two effects on the oscillations. On the one hand, the spreading of the wave packet compensates for the spatial separation of the mass states, hence restoring parts of their interferences. On the other hand, dispersion reduces the overlapping fraction of the wave packets, and thus the interference or oscillation effects cannot be fully restored. 
Moreover, it also modifies the flavor oscillation phases:
\begin{equation}\label{def_phi}
\phi_{ij} \equiv \dfrac{2\pi L}{L^\mathrm{osc}_{ij}} + \left(\frac{1}{2} \mbox{tan}^{-1} (y_{ij}) - \lambda_{ij}y_{ij} \right),
\end{equation}
with deviations from the standard plane-wave oscillation phase written in the parentheses. If $y_{ij}$ = 0, then $\phi_{ij}$ just reduce to the standard plane-wave oscillation phases.

\subsection{Decoherence effect due to delocalization}\label{sec2.3}
The decoherence effect mentioned in the previous subsection is due to the separation of different neutrino wave packets. With larger values of $\sigma_\mathrm{wp}$, the corresponding decoherence effect would be more significant. On the other hand, there also exists another kind of decoherence effect not described in the previous subsections, due to the delocalization of the production and detection processes. Different with what we have studied above, the decoherence effect from delocalization\footnote{In the following content, we will call this kind of decoherence effect as delocalization effect in order to separate it from the decoherence effect due to separation of wave packets.} will become significant only when $\sigma_\mathrm{wp}$ is extremely small. 

\begin{figure}
\centering
 \includegraphics[scale=0.32]{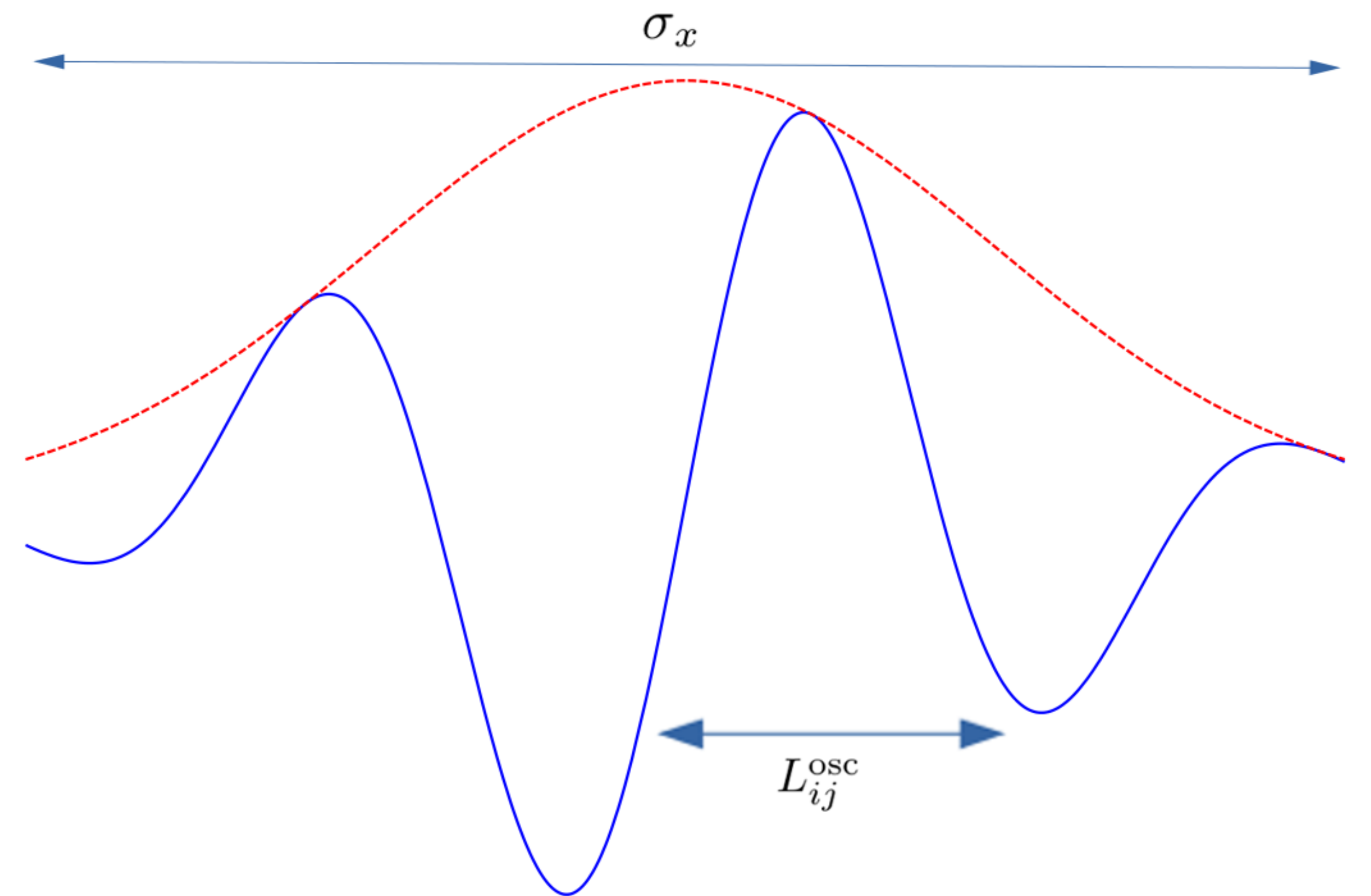}
\caption{Schematic representation of the decoherence effects due to the delocalization term in Eq. \ref{Pee_complete}. When the width of wave packet is comparable or even larger than the oscillation length, the oscillation would be destroyed.}
\label{fig:localization}
\end{figure}

In reality, we have assumed the terms $\propto \dfrac{\frac{(\Delta m_{ij}^2)^2}{E^4}}{\sigma_\mathrm{wp}^2}$ are negligible in the previous subsections since they are inversely proportional to $\sigma_\mathrm{wp}^2$. However, if $\sigma_\mathrm{wp}$ is extremely small, which means that the spatial width of the wave packet $\sigma_x$ is large, it will lead to the other kind of damping signature of the neutrino oscillations. With these delocalization terms taken into account, a more complete $\bar\nu_e$ survival probability is given by\footnote{The details of the derivation of oscillation formula can be found in reference \cite{Chan:2015mca}.}:


\begin{align}\label{Pee_complete}
 P_{\nu_\alpha \rightarrow \nu_\beta}(L) & \approx \sum_{ij} \left\{ U_{\alpha i}^\ast U_{\beta i}U_{\alpha j}U_{\beta j}^\ast \mathrm{exp}\left[-i\dfrac{\Delta m^2_{ij} L}{2E}\right] \right\} \notag \\
 & \left\{ \left(\dfrac{1}{1+y_{ij}^2}\right)^{\frac{1}{4}}\mathrm{exp}(-\lambda_{ij})\mathrm{exp}\left(\frac{-i}{2} \mbox{tan}^{-1} (y_{ij})\right) \mathrm{exp}(i\lambda_{ij}y_{ij}) \right\} \mathrm{exp}(-\gamma_{ij}), \\
                                       \mbox{where } & \gamma_{ij} = \dfrac{\dfrac{1}{16}\dfrac{(\Delta m_{ij}^2)^2}{E^4}}{1+y_{ij}^2}\cdot\dfrac{1}{\sigma_\mathrm{wp}^2} = \dfrac{\pi^2}{(1+y_{ij}^2)}\cdot\dfrac{\sigma_x^2}{(L^\mathrm{osc}_{ij})^2}, \notag
\end{align}
The additional damping factor exp(-$\gamma_{ij}$) is important when $\sigma_x$ becomes comparable to $L^\mathrm{osc}_{ij}$. Fig. \ref{fig:localization} further describes the extreme case when $\sigma_x$ $\gg$ $L^\mathrm{osc}_{ij}$; in such circumstances it is difficult to observe the oscillation effect. 

In fact, in neutrino oscillation, one of the coherence conditions is that the intrinsic production (and also detection) energy uncertainties are much larger than the energy difference between different mass eigenstates ($\Delta E_{ij}$) \cite{Akhmedov:2012uu}, namely,
\begin{equation}\label{condition}
 \Delta E_{ij} \equiv E_i - E_j \sim \dfrac{\Delta m_{ij}^2}{E_\nu} \ll \sigma_\nu \equiv E_\nu\sigma_\mathrm{wp},
\end{equation}
Eq. (\ref{condition}) implies that in order to measure the interferences between different mass eigenstates, the spatial uncertainty $\sigma_x$ has to be much smaller than the oscillation length. Namely, $\sigma_x \ll L_\mathrm{osc}$.


The condition in Eq. (\ref{condition}) is satisfied in most reactor neutrino oscillation measurements, since the production and detection processes are (spatially) localized in regions smaller than the reactor and detector sizes, which are much smaller than the oscillation length\footnote{However, in the measurement of sterile neutrino oscillation, the oscillation length is expected to be short and may be comparable to the spatial uncertainty $\sigma_x$. In this case the decoherence effect coming from the localization term should not be neglected.}. Therefore, in most of the neutrino oscillation experiments, 
\begin{align}\label{lower_approx}
 \gamma_{ij} = \dfrac{\pi^2}{(1+y_{ij}^2)}\cdot\dfrac{\sigma_x^2}{(L^\mathrm{osc}_{ij})^2} & \approx 0, 
\end{align}
Thus, in most circumstances, the delocalization damping term $\mathrm{exp}(-\gamma_{ij})$ can be safely neglected. 




It is obvious that in Eq. (\ref{Pee_complete}), the terms of $\gamma_{ij}$ do not allow $\sigma_\mathrm{wp}$ to go to zero as $\gamma_{ij} \propto 1/ \sigma_\mathrm{wp}$. The decoherence effect due to delocalization would lead to a lower bound of the possible range of $\sigma_\mathrm{wp}$. On the other hand, the terms of $\lambda_{ij}$ and $y_{ij}$ are significant only when $\sigma_\mathrm{wp}$ is large. This implies that if $\sigma_\mathrm{wp}$ is extremely small, the decoherence due to spatial separations is negligible and even dispersion effect is also subdominant. $\lambda_{ij}$ and $y_{ij}$ would lead to the upper limit of the allowed region of $\sigma_\mathrm{wp}$. In a word, in the wave-packet treatment, only one extra parameter is introduced ($\sigma_\mathrm{wp}$ or $\sigma_\nu$) but it can describe two different decoherence effects. It is because the energy uncertainty ($\sigma_\mathrm{wp}$) being either too large or too small\footnote{Small energy uncertainty implies large spatial uncertainty.} would also destroy the oscillation. Both the separation of wave packets and the delocalization effect depend on the initial width of the neutrino wave packet.

\subsection{The estimation of $\sigma_\mathrm{wp}$}\label{sec2.4}
The value of this parameter or the size of neutrino wave packet has not come to a strong conclusion yet. Up to now, Daya Bay is the only neutrino experiment which provides the experimental constraints on this parameter. On the other hand, there have been different theoretical estimations of the sizes of neutrino wave packets produced in different situations. For example, Reference \cite{Hernandez:2011rs} uses the pion decay length to estimate the width of neutrino wave packet in the MINOS experiment and argues that there could be significant decoherence effect in the active to sterile neutrino oscillation in MINOS\footnote{Please notice that the decoherence effect in Reference \cite{Hernandez:2011rs} is due to the delocalization, different with the decoherence effect due to separations of wave packets but it could also destroy the oscillation.}. Regarding to reactor neutrino experiments, Reference \cite{SigmaxRange} provides an estimation based on the mean free path and mean thermal velocity of the production process and suggests that $\sigma_x$ $\sim$ $10^{-6}$ m, which implies that $\sigma_\mathrm{wp}$ $\sim$ $10^{-7}$ in reactor neutrino experiments. Meanwhile, References \cite{Rich:1993wu,Grimus:1996av} suggest that the neutrino emission process is expected to be localized at the scale of inter-atomic distance, and so $\sigma_x$ $\lesssim$ $10^{-10}$ m, implying $\sigma_\mathrm{wp}$ $\sim$ $10^{-3}$ or even larger. If on the other hand, one takes the uncertainty of a nucleon's position in a nucleus as $\sigma_x$, then $\sigma_\mathrm{wp}$ could be much larger, even of order 1. The estimated sizes of neutrino wave packets from different approaches could be different by a few orders of magnitude. Moreover, as pointed out by Reference \cite{Beuthe:2001rc}, the relation between the decay time of the source and the wave packet size of the oscillating particle is not direct. The decay time only puts an upper bound on the wave packet length. There is still no experimental support for such an assumption. In this paper we do not calculate or suggest the theoretical value of $\sigma_\mathrm{wp}$.

\section{The constraints from medium-baseline reactor neutrino oscillation experiment}\label{sec3}
Thus far, there are no significant signals of decoherence and dispersion effects caused by the neutrino wave-packet treatment. The Daya Bay collaboration have analyzed their data with the neutrino wave-packet framework and managed to provide lower and upper limits on the value of $\sigma_\mathrm{wp}$:\footnote{Please note that our definition of $\sigma_\nu$ is different with the corresponding parameter $\sigma_p$ in reference \cite{An:2016pvi}: $\sigma_\nu$ = $\sqrt{2} \sigma_p$. Therefore the value of $\sigma_\mathrm{wp}$ is also $\sqrt{2}$ times larger than the $\sigma_\text{rel}$ in that paper.} $2.38 \times 10^{-17} < \sigma_\text{rel} < 0.23$ at 95\% C.L. \cite{An:2016pvi}.

In this paper, we focus on the analysis of wave-packet impact in the future MBRO experiment(s) only. Hence we are just interested in the electron anti-neutrino survival probability, which is given by:

\begin{align}\label{Pee_app}
 P_{\bar{e}\bar{e}} = 1 - & \frac{1}{2}\mathrm{cos}^4(\theta_{13})\mathrm{sin}^2(2\theta_{12})[1 - (\dfrac{1}{1+y_{21}^2})^{\frac{1}{4}}\mathrm{exp}(-\lambda_{21})\mathrm{exp}(-\gamma_{21})\mathrm{cos}(\phi_{21})] - \notag \\
          & \frac{1}{2}\mathrm{sin}^2(2\theta_{13})\mathrm{cos}^2(\theta_{12})[1 - (\dfrac{1}{1+y_{31}^2})^{\frac{1}{4}}\mathrm{exp}(-\lambda_{31})\mathrm{exp}(-\gamma_{31})\mathrm{cos}(\phi_{31})] - \notag \\
          & \frac{1}{2}\mathrm{sin}^2(2\theta_{13})\mathrm{sin}^2(\theta_{12})[1 - (\dfrac{1}{1+y_{32}^2})^{\frac{1}{4}}\mathrm{exp}(-\lambda_{32})\mathrm{exp}(-\gamma_{32})\mathrm{cos}(\phi_{32})].
\end{align}

We use Eq. (\ref{Pee_app}) to perform numerical simulations and estimate the sensitivities of future medium baseline reactor experiment(s) on the constraints of $\sigma_\mathrm{wp}$. The oscillation parameter values are taken from global analysis \cite{Tanabashi:2018oca} as $\Delta m^2_{21}$ = 7.53 $\times$ 10$^{-5}$ eV$^2$, ($\Delta m^2_{31}$ + $\Delta m^2_{32}$) / 2 = 2.548 $\times$ 10$^{-3}$ eV$^2$, sin$^2\theta_{12}$ = 0.307 and sin$^2\theta_{13}$ = 0.0212. We quantify the sensitivity of $\sigma_\mathrm{wp}$ by employing the least-squares method, based on a $\chi^2$ function given by
\begin{align}\label{chi2}
 \chi^2 & = \sum_i^{N_\mathrm{bin}} \dfrac{[T_i - F_i(1 + \eta_R + \eta_d + \eta_i)]^2}{T_i} + (\dfrac{\eta_R}{\sigma_R})^2 + (\dfrac{\eta_d}{\sigma_d})^2 + \sum_i^{N_\mathrm{bin}}(\dfrac{\eta_i}{\sigma_{s,i}})^2
\end{align}
where $T_i$ is measured neutrino event in the $i$th energy bin, and $F_i$ is the predicted number of neutrino events with oscillations taken into account (the fitting event rate). $\eta$ with different subscripts are nuisance parameters corresponding to reactor-related uncertainty ($\sigma_R$), detector-related uncertainty ($\sigma_d$) and shape uncertainty ($\sigma_s$). According to the References \cite{Yifang, An:2015jdp, Wang:2016vua}, $\sigma_R$, $\sigma_d$ and $\sigma_{s,i}$ are assumed to be 2\%, 1\% and 1\% at MBRO experiment(s), respectively. 

\subsection{The upper and lower limits}\label{sec3.1}
\begin{figure*}[!htbp]
\centering
 \includegraphics[scale=0.53]{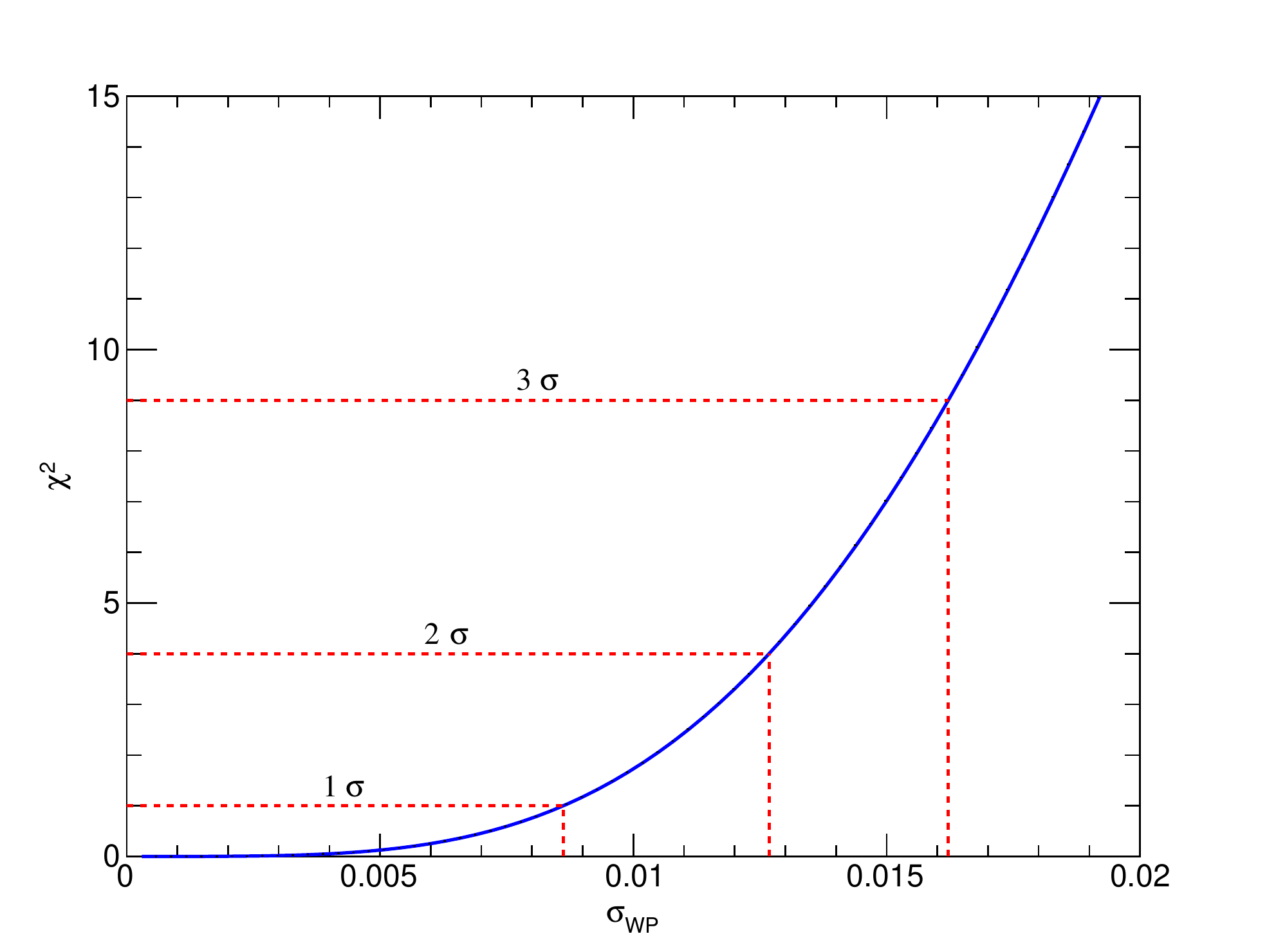} \\ \includegraphics[scale=0.53]{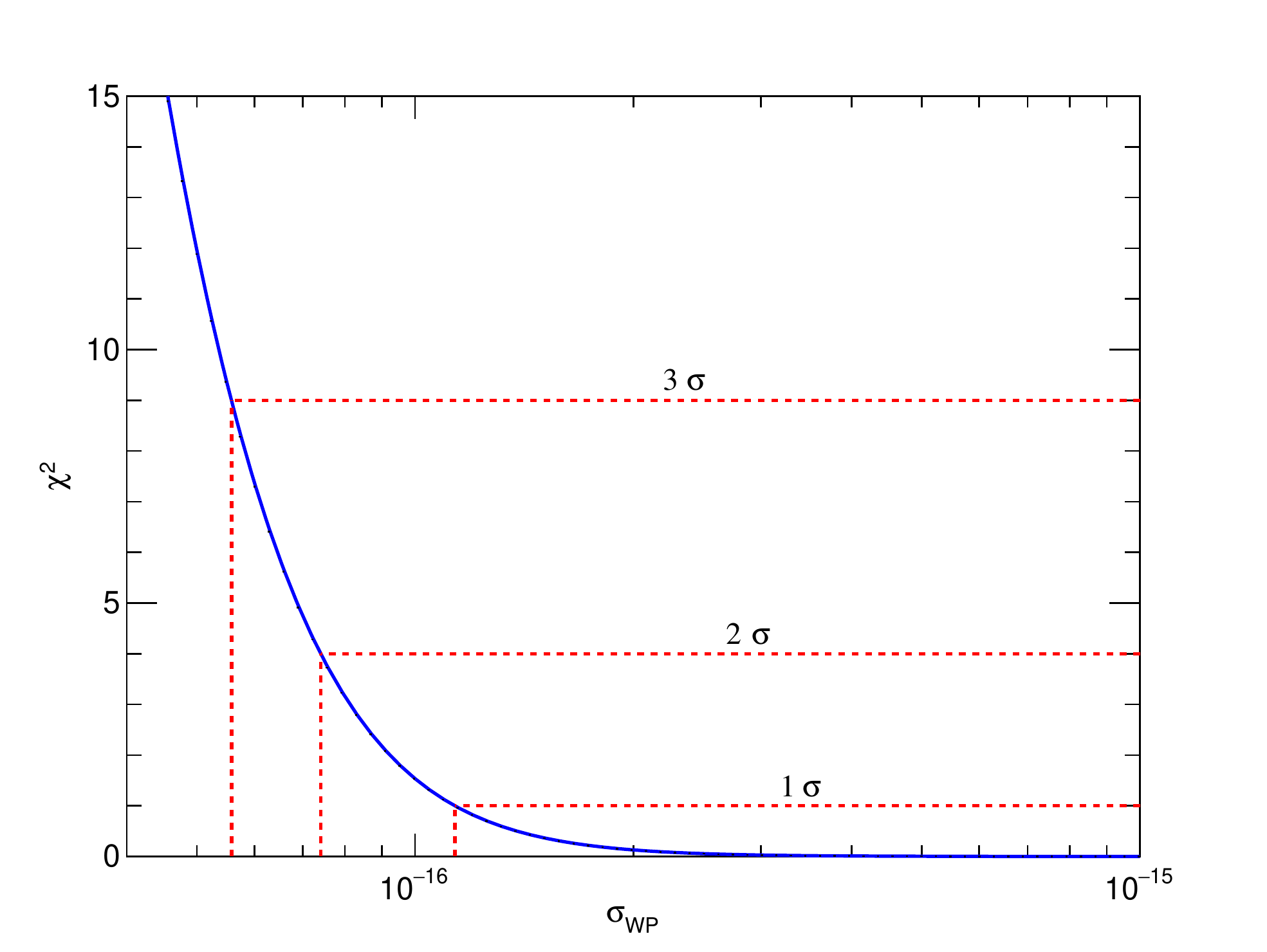} 
\caption{Top: the 1, 2, 3 $\sigma$ C.L upper bound on $\sigma_\mathrm{wp}$ at the future medium-baseline reactor neutrino oscillation experiment(s). Bottom: the 1, 2, 3 $\sigma$ C.L lower bound on $\sigma_\mathrm{wp}$.}
\label{fig:sigma_wp_sensitivity}
\end{figure*}

If $\sigma_\mathrm{wp}$ is relatively large, in this case the decoherence effect is resulted from the separations of wave packets. Moreover, in this region, the dispersion effect could be significant and lead to modifications on the decoherence effect \cite{Chan:2015mca}, which makes the oscillation patterns more complicated. The future MBRO experiment is expected to be sensitive to the decoherence and dispersion effects and provide constraint on decoherence parameter $\sigma_\mathrm{wp}$. The results of our simulations are correspondingly shown in the top panel of Fig. \ref{fig:sigma_wp_sensitivity}. The resulting upper bounds are found to be 0.0086, 0.0127, 0.0162 at 1, 2, 3 $\sigma$ C.L, respectively. 

On the other hand, if $\sigma_\mathrm{wp}$ is extremely small, which means that the spatial uncertainty of the neutrino wave packet is large, the delocalization will lead to the other kind of decoherence effect as the damping factor exp(-$\gamma_{ij}$) in Eq. (\ref{Pee_app}). However, in the region of extremely small $\sigma_\mathrm{wp}$, the dispersion effect can be safely neglected. The results of the lower limit based on our numerical simulations is shown in the bottom panel of Fig. \ref{fig:sigma_wp_sensitivity}. The lower bounds are found to be 1.13 $\times$ $10^{-16}$, 7.42 $\times$ $10^{-17}$, 5.58 $\times$ $10^{-17}$ at 1, 2, 3 $\sigma$ C.L.
As discussed in the subsection \ref{sec2.3}, the decoherence effect due to delocalization is significant only when the spatial width of the neutrino wave packet ($\sigma_x$) is comparable to the oscillation length ($L^\mathrm{osc}$). Since the MBRO experiment(s) is expected to be able to observe both the atmospheric- and solar- $\Delta m^2$ driven oscillations, it is supposed to be sensitive to the delocalization effect in a large range. However, $L^\mathrm{osc}_{32}$ $\sim$ $O(1 \mathrm{km})$, $L^\mathrm{osc}_{21}$ $\sim$ $O(50 \mathrm{km})$. It means that $\sigma_x$ has to be around a few hundred meters, otherwise the delocalization effect will be insignificant in the future MBRO experiment(s). 

In fact, Fig. \ref{fig:sigma_wp_sensitivity} shows that the lower limit of our simulation is around $\sigma_\mathrm{wp} \sim O(10^{-16})$, which corresponds to $\sigma_x$ $\sim$ $O(1 \mathrm{km})$. Nevertheless, it is obvious that the spatial width of the neutrino wave packet should not be larger than the dimensions of the reactor cores and detectors, which are just around $O(10)$ meters. Therefore, we conclude that the future MBRO experiment(s) cannot provide a stringent lower bound to $\sigma_\mathrm{wp}$. In the following sections, we will focus on the studies of upper bound and neglect the decoherence effect due to delocalization. 

\subsection{The impacts of statistical, shape uncertainties and detector energy resolutions on the sensitivity}\label{sec3.2}
\begin{figure*}[!htbp]
\centering
 \includegraphics[scale=0.53]{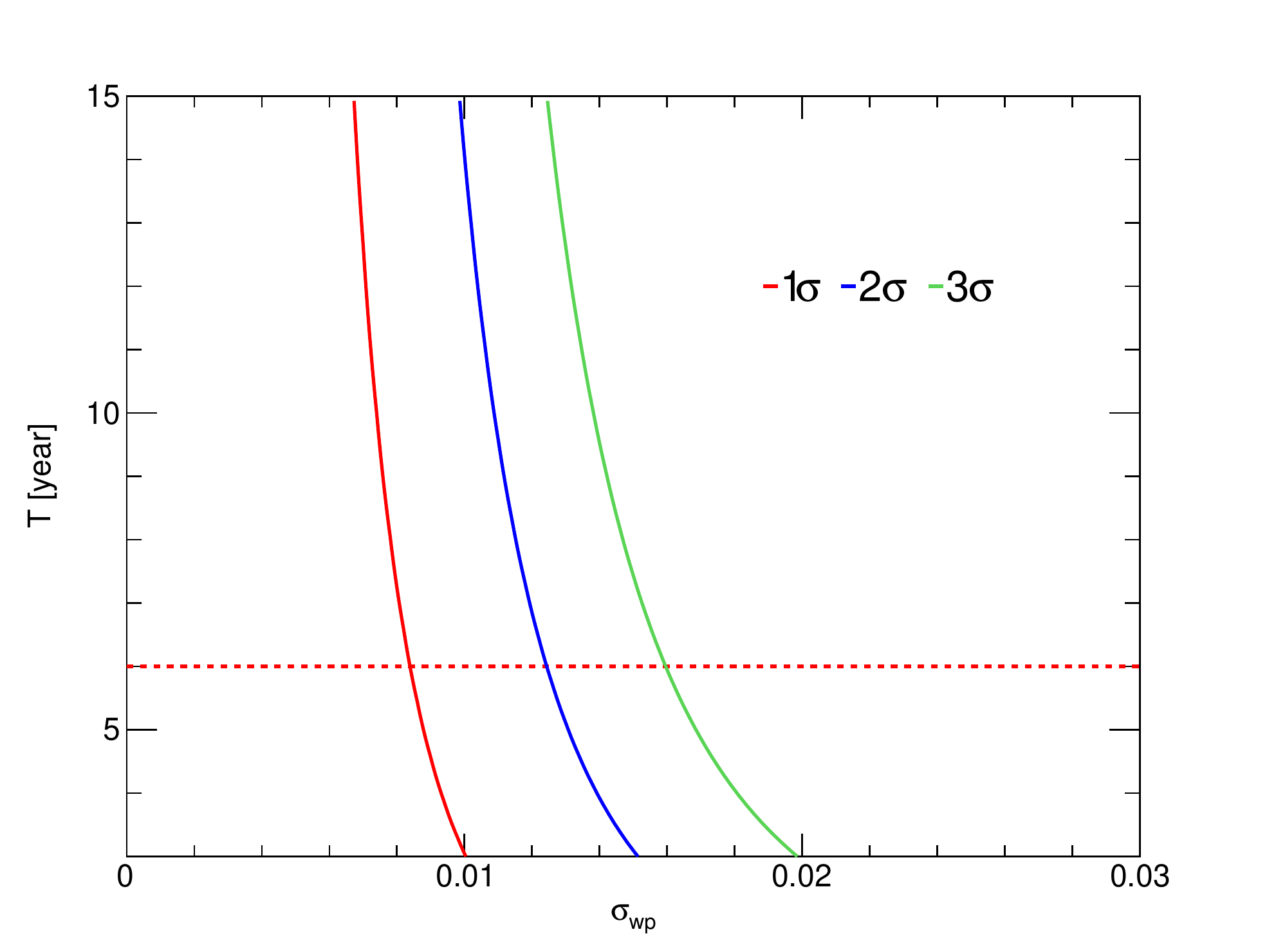} 
\caption{The sensitivity of constraining $\sigma_\mathrm{wp}$ vs. years of data taking in the future medium-baseline detector. The horizontal red dashed line represents the nominal running time (six years) proposed in reference \cite{An:2015jdp}.}
\label{fig:constrain_sigma_wp_events_FD_only}
\end{figure*}

The sensitivities in Fig. \ref{fig:sigma_wp_sensitivity} corresponds to the assumption of 6 years data-taking and 1\% shape uncertainties. In this subsection, we examine whether reducing such uncertainties can significantly improve the sensitivity of constraining $\sigma_\mathrm{wp}$. Fig. \ref{fig:constrain_sigma_wp_events_FD_only} shows the effect of statistics on the study of decoherence effect. The y-axis corresponds to the time of data taking. Nevertheless, our simulations suggest that after data collecting for more than 10 years, the limits of $\sigma_\mathrm{wp}$ is barely improved by collecting more oscillation events.

\begin{figure*}[!htbp]
\centering
 \includegraphics[scale=0.53]{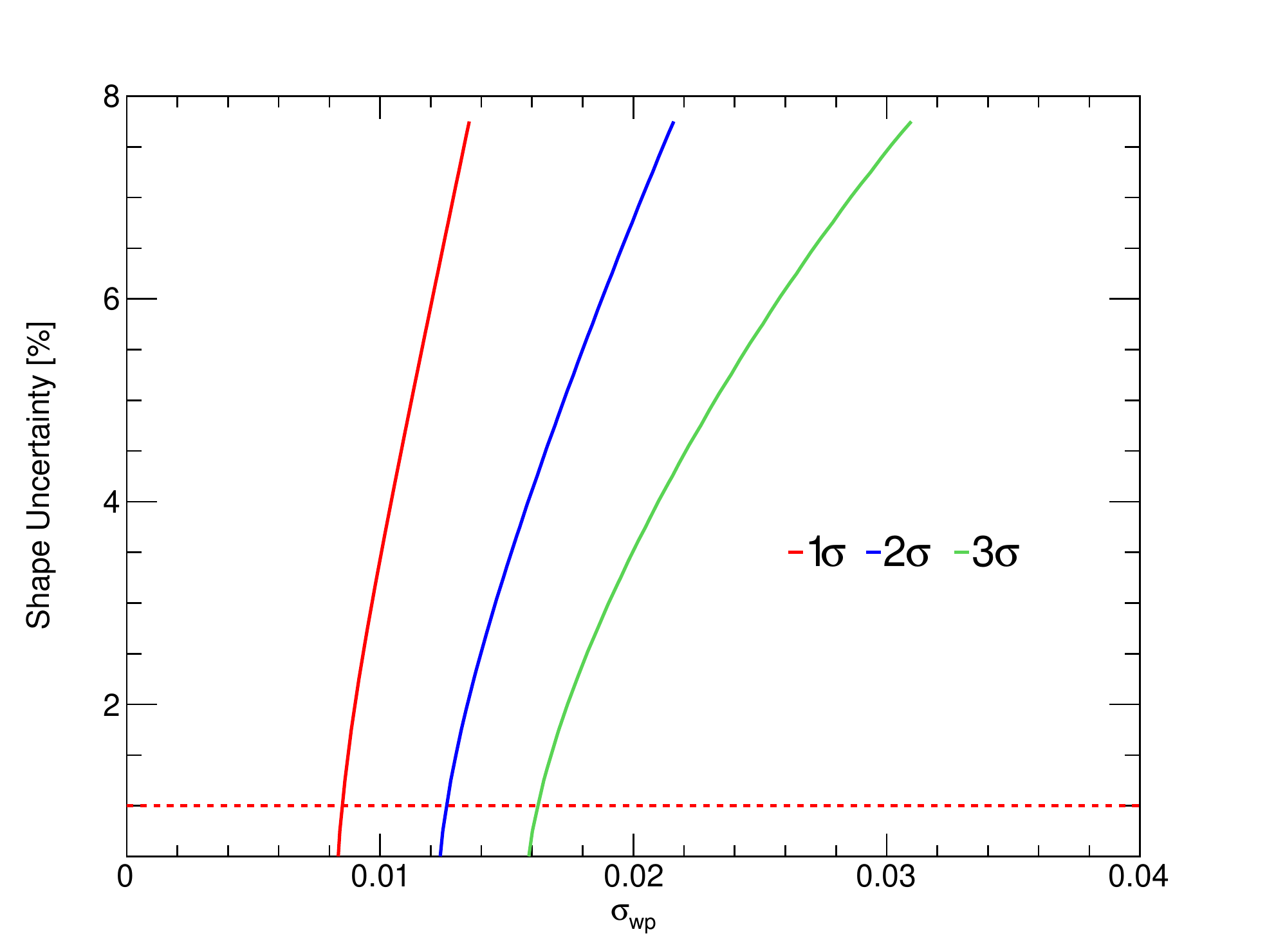} 
\caption{The sensitivity of constraining $\sigma_\mathrm{wp}$ vs. shape uncertainties in the future medium-baseline detector. The horizontal red dashed line represents the suggested shape uncertainty (1\%) in references \cite{An:2015jdp, Yifang, Wang:2016vua}.}
\label{fig:constrain_sigma_wp_ShapeEr_FD_only}
\end{figure*}

On the other hand, it is also worthwhile to examine the impact of systematic uncertainties on the study of neutrino wave-packet treatment. Since the wave-packet impact modifies the neutrino oscillation pattern, we believe that the shape uncertainty is the most important uncertainties in the future MBRO experiment(s). Conventionally, the shape uncertainty is assumed to be 1\% for all energy bins \cite{Yifang,An:2015jdp,Wang:2016vua}. Nevertheless, recently there are literatures suggest that the shape uncertainties could be underestimated. Compared with the conventional prediction, the measured IBD positron (antineutrino) energy spectrum from Daya Bay, Reno and Double Chooz \cite{An:2012eh, Ahn:2012nd, Abe:2011fz} show an event excess in in the region of 4 to 6 MeV prompt energy.
Moreover, the precise shape of the flux spectrum is hard to be determined, which could lead to fine structure and additional shape uncertainties in the analyses of the future MBRO experiment(s) \cite{Dwyer:2014eka,Sonzogni:2017voo,Capozzi:2015bpa,Cheng:2020ivh}. However, we just assume the conventional 1$\%$ shape uncertainty in Figs. \ref{fig:sigma_wp_sensitivity} and \ref{fig:constrain_sigma_wp_events_FD_only}, which is samed with the References \cite{An:2015jdp,Yifang,Wang:2016vua}. To further investigate the impact of shape uncertainty, we alter its values and show the resulting constraint on $\sigma_\mathrm{wp}$ in Fig. \ref{fig:constrain_sigma_wp_ShapeEr_FD_only}.

Fig. \ref{fig:constrain_sigma_wp_ShapeEr_FD_only} shows that the upper bounds on $\sigma_\mathrm{wp}$ become weaker if shape uncertainty increases, since the observation of decoherence effect or any damping signature depends on the shape analysis. However, the impact of shape uncertainty is not large at all, because as long as the medium-baseline detector manages to resolve multiple neutrino oscillations, the decoherence effect can still be strongly constrained even if the uncertainties of each energy bin become larger. Moreover, our results also suggest that the unknown shape of reactor neutrino flux or issues of potential fine structure \cite{Dwyer:2014eka,Sonzogni:2017voo,Capozzi:2015bpa,OurPaper1} would not significantly affect the study of neutrino wave-packet impact.

\begin{figure*}[!htbp]
\centering
 \includegraphics[scale=0.53]{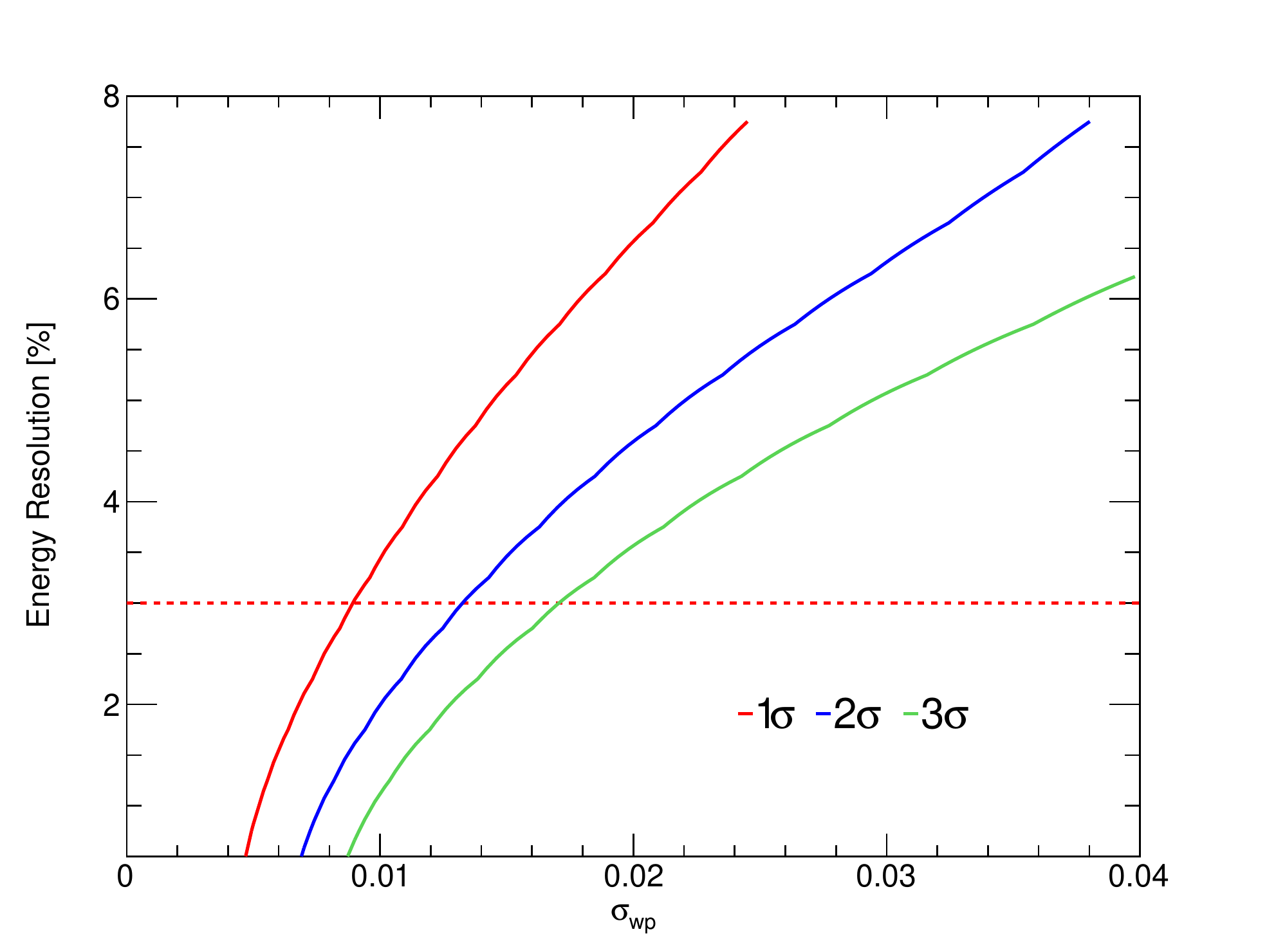} 
\caption{The sensitivity of constraining $\sigma_\mathrm{wp}$ vs. detector energy resolution. The horizontal red dashed line represents the proposed (3$\%$) energy resolution of the future MBRO experiment(s) \cite{An:2015jdp}.}
\label{fig:constrain_sigma_wp_Res_FD_only}
\end{figure*}

Besides, the detector energy resolution is believed to be more crucial in studying the potential decoherence effect. Fig. \ref{fig:constrain_sigma_wp_Res_FD_only} shows the importance of detector energy resolution in constraining the parameter $\sigma_\mathrm{wp}$. The red dot-dashed line in this figure represents the results from the proposed detector energy resolution at the future MBRO experiment(s). As a result, the 3 $\sigma$ C.L upper bound on $\sigma_\mathrm{wp}$ is found to be around 0.0162, which is samed with the previous result in Fig. \ref{fig:sigma_wp_sensitivity}.

Comparing Fig. \ref{fig:constrain_sigma_wp_Res_FD_only} with Figs. \ref{fig:constrain_sigma_wp_events_FD_only} and \ref{fig:constrain_sigma_wp_ShapeEr_FD_only}, we find that the future MBRO experiment(s) provides an ideal platform for the study on neutrino wave-packet hypothesis and gives rise to the fine upper bounds on the decoherence effect because of the unprecedented detector energy resolution. In addition, improvements on the statistical and systematic uncertainties could further improve the sensitivity, but the effects are not sizable.

\subsection{Comparing with the current constraints from Daya Bay}\label{sec3.3}
As shown in Fig. \ref{fig:sigma_wp_sensitivity}, the 1, 2, 3 $\sigma$ C.L upper bounds on $\sigma_\mathrm{wp}$ from the future MBRO experiment(s) are 0.0086, 0.0127, 0.0162 respectively. It means that the MBRO experiment(s) can constrain $\sigma_\mathrm{wp}$ at the order of $O(10^{-2})$, which is around 10 times better than the current results from Daya Bay, thanks to the length of baseline and excellent detector energy resolution of the future MBRO experiment(s). Daya Bay data put an upper limit: $\sigma_\mathrm{rel} < 0.2$ at 95\% C.L.\footnote{Please keep in mind that due to different definition of the width of neutrino wave packet, our $\sigma_\mathrm{wp}$ is different with the $\sigma_\text{rel}$ in the Daya Bay paper \cite{An:2016pvi}: $\sigma_\mathrm{wp}$ = $\sqrt{2} \sigma_\text{rel}$. Converting to our definition, the Daya Bay 95\% C.L. upper bound corresponds to $\sigma_\mathrm{wp} < 0.283$.} \cite{An:2016pvi}, while our simulations suggest that the future MBRO experiment(s) can improve it to $\sigma_\mathrm{wp} < $ 0.0125 at 95\% C.L.

On the other hand, regarding to the delocalization effect, the Daya Bay 95\% C.L. lower limit is given by:
\begin{align}\label{DB_lower_limit}
 \sigma_\mathrm{rel} > 2.38\times 10^{-17}
\end{align}
Convert $\sigma_\mathrm{rel}$ to our definition:  
\begin{align}\label{DB_lower_limit_WP}
 \sigma_\mathrm{wp} > 3.37 \times 10^{-17}
\end{align}
Our simulations suggest that the 1, 2, 3 $\sigma$ C.L lower bounds on $\sigma_\mathrm{wp}$ from the future MBRO experiment(s) are $1.13\times 10^{-16}, 7.42\times 10^{-17}, 5.58\times 10^{-17}$ respectively. The 95\% C.L. lower limit is given by:
\begin{equation}\label{Our_95CL_lower_limit}
  \sigma_\mathrm{wp} > 7.51 \times 10^{-17}.
\end{equation}

Our lower limit is larger than the Daya Bay published result, which means that the lower limit of future MBRO experiment(s) is also better than the one of Daya Bay, but the difference is not large. Nevertheless, the practical constraint of $\sigma_x$ $\lesssim$ $O(10)$ m corresponds to $\sigma_\mathrm{wp}$ $\gtrsim$ $O(10^{-14})$. It implies that both the lower bounds from current and future reactor neutrino experiments are actually weaker than the obvious constraints based on consideration of sizes of reactor cores and detectors. As mentioned before, it is because the delocalization effect is significant only when $\sigma_x$ $\approx$ $L^\mathrm{osc}$. Therefore, it is expected to be significant only in the oscillations corresponding to $\Delta m^2$ $\sim$ 0.1 eV$^2$.

\section{The potential impact on the precision measurement of $\theta_{12}$ and other oscillation parameters}\label{sec4}
If the decoherence and dispersion effects are signifcant in MBRO experiment, they could give rise to modification of oscillation patterns and thus affect the identification of the neutrino mass ordering, and also the precision measurements of oscillation parameters. The potential wave-packet impact on the resolution of neutrino mass hierarchy in MBRO experiment can be found in reference \cite{Chan:2015mca}. In this section, we will discuss the wave-packet impact on the measurements of oscillation parameters.

Besides studying the neutrino mass hierarchy and observations of multiple oscillation cycles, the future MBRO experiment(s) is also expected to provide the unprecedented precision measurements of $\theta_{12}$, $\Delta m^2_{21}$ and $|\Delta m^2_{ee}|$ to better than 1\% \cite{An:2015jdp, Cao:2017drk}. Nevertheless, the neutrino wave-packet treatment could potentially lead to the biases on these precision measurements. In order to explore the wave-packet impacts on the future precision measurements, we use Eq. (\ref{Pee_app}), and set $\sigma_\mathrm{wp}$ and also other oscillation parameters\footnote{According to Reference \cite{Chan:2015mca}, the decoherence effect could affect the measurements of the oscillation parameters such as the mixing angles and mass square differences. Nevertheless, according to the analysis from Reference \cite{An:2016pvi}, the measurement of $\theta_{13}$ from Daya Bay is barely affected by the wave-packet impact and the value of sin$^2 2\theta_{13}$ is not changed. Since Daya Bay is believed to provide the most precise measurement on $\theta_{13}$, in our simulation we use the value of sin$^2 2\theta_{13}$ as Daya Bay reported and set $\theta_{13}$ as a fixed parameter.} as free parameters in our simulations.
\begin{figure*}[!htbp]
\centering
 \includegraphics[scale=0.56]{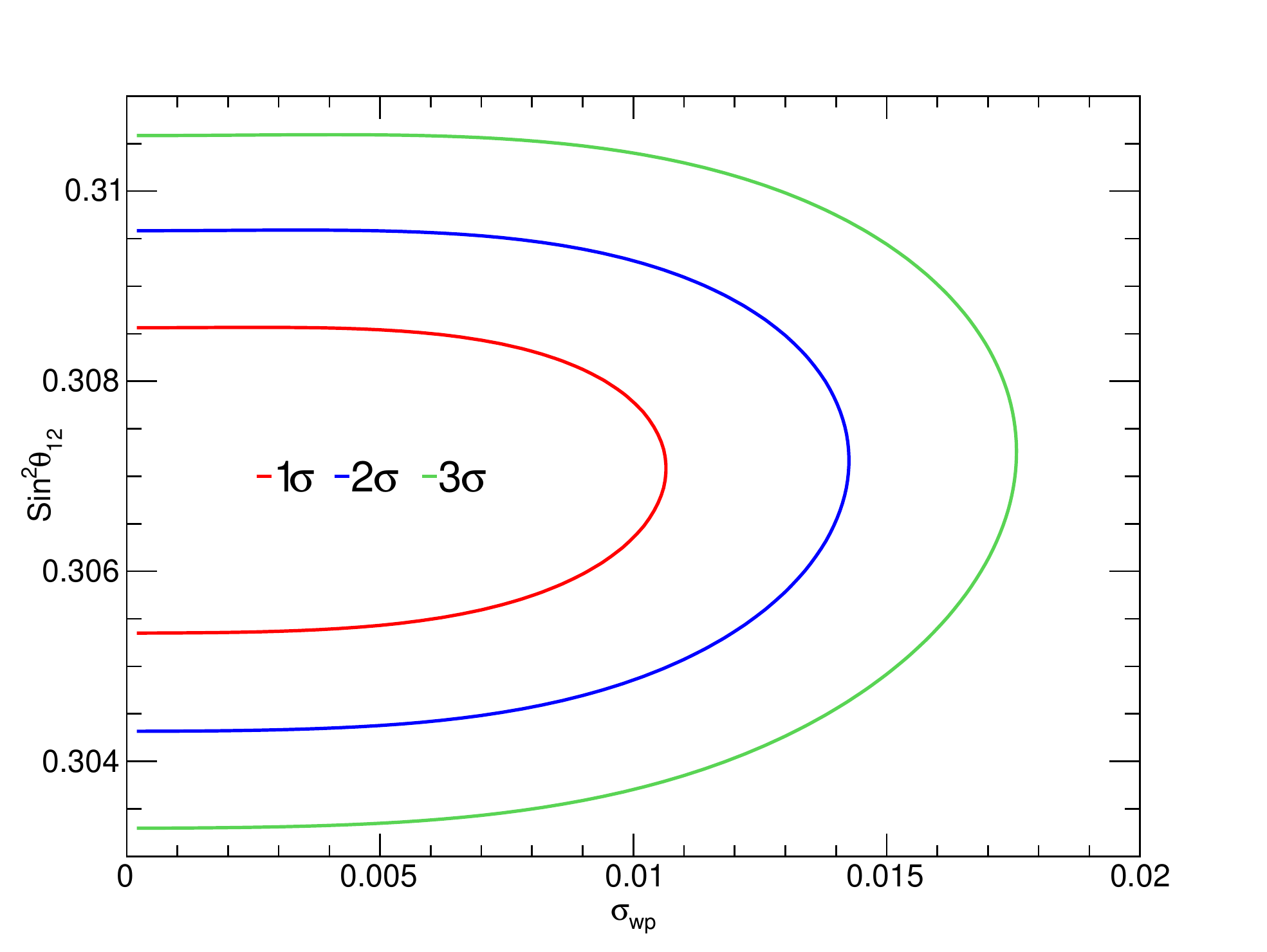} 
\caption{The 1 $\sigma$ (red), 2 $\sigma$ (blue) and 3 $\sigma$ (green) constraints on ``$\sigma_\mathrm{wp}$ vs sin$^2 \theta_{12}$'' for the large $\sigma_\mathrm{wp}$ region.}
\label{fig:constrain_sigma_wp_up_s212}
\end{figure*}

\begin{figure*}[!htbp]
\centering
 \includegraphics[scale=0.4]{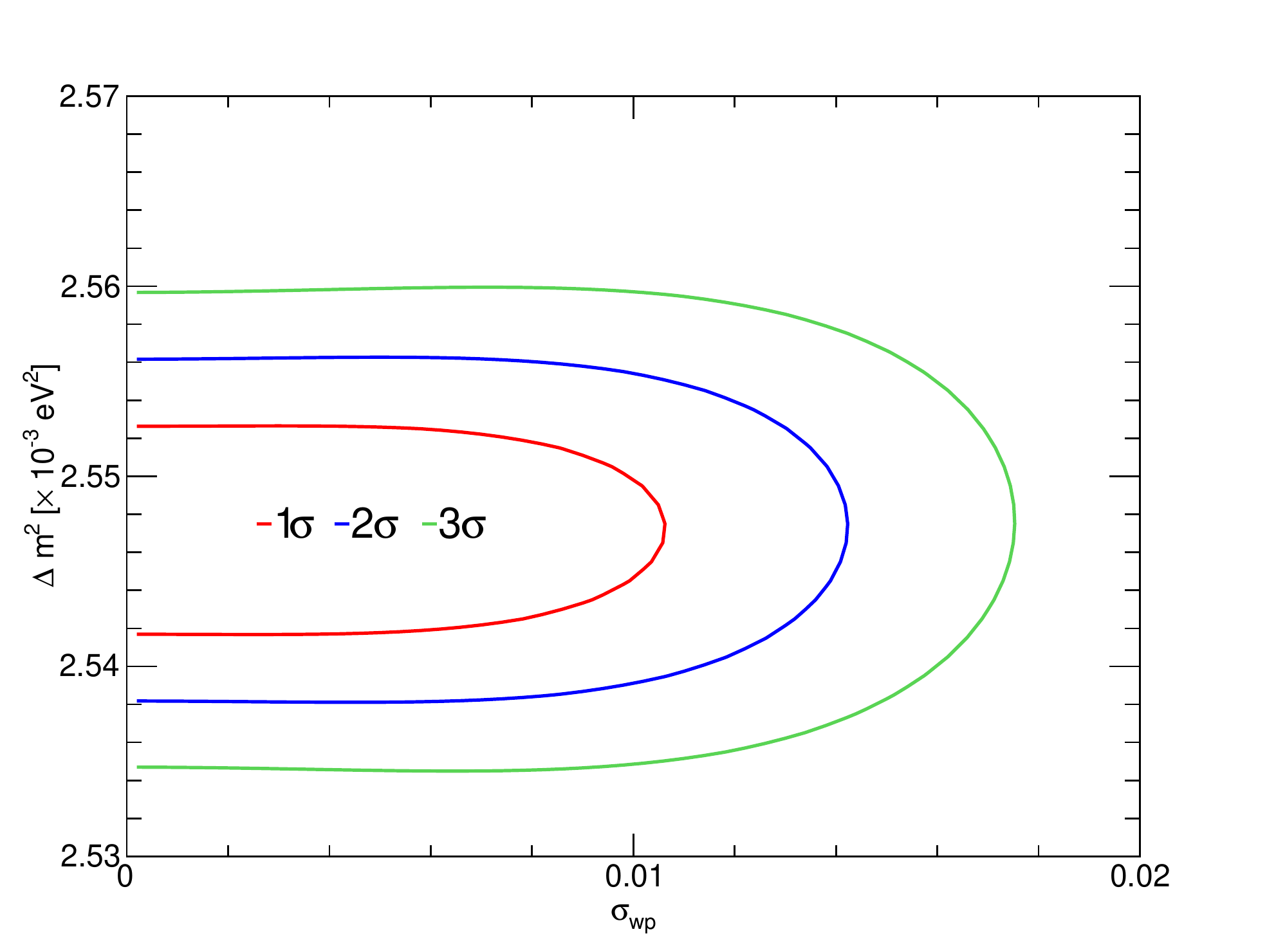} \quad \includegraphics[scale=0.4]{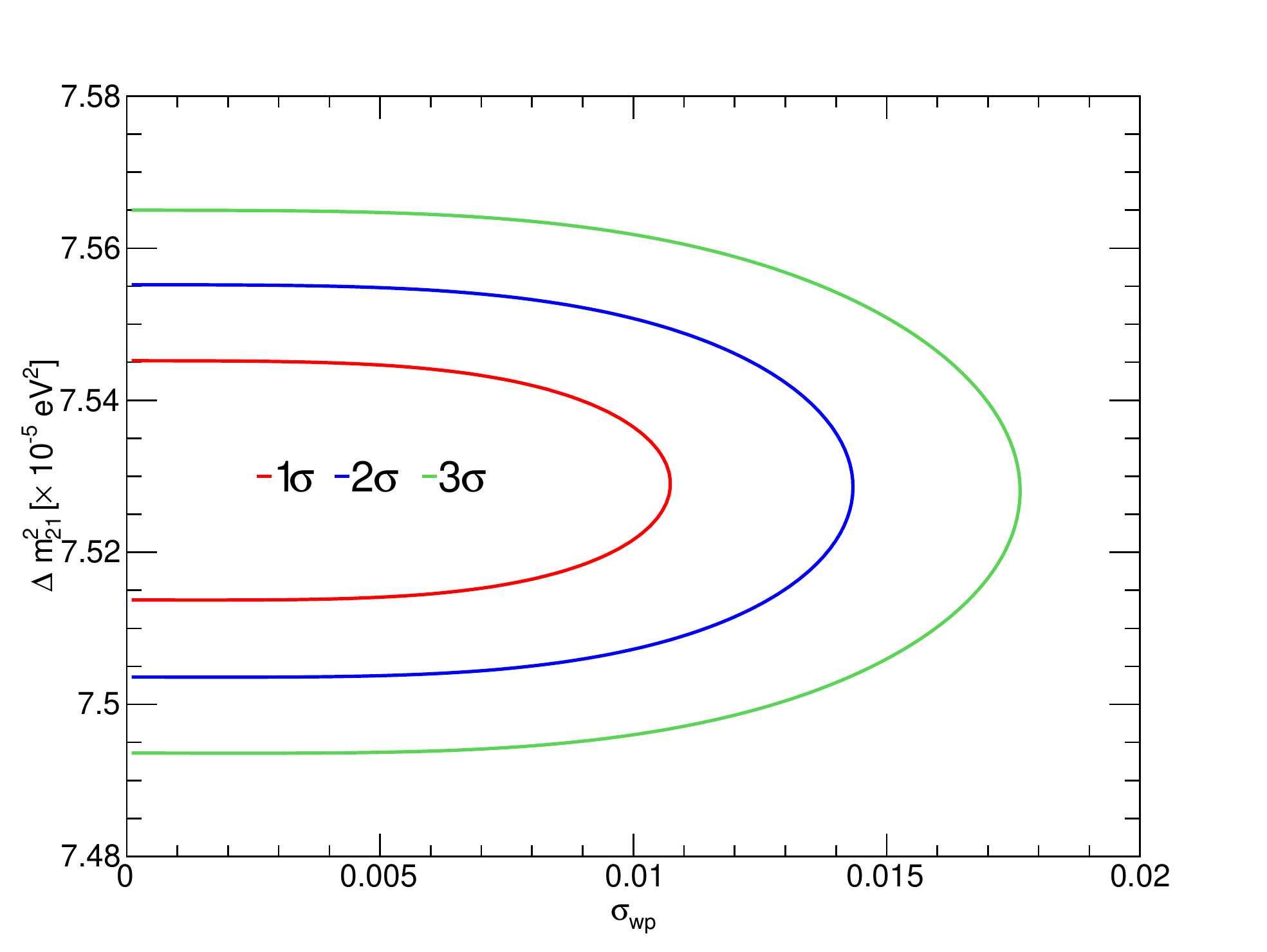}
\caption{Left: The 1 $\sigma$ (red), 2 $\sigma$ (blue) and 3 $\sigma$ (green) constraints on ``$\sigma_\mathrm{wp}$ vs $\Delta m^2$'' for the large $\sigma_\mathrm{wp}$ region.\\
Right: Same as the left panel, but for ``$\sigma_\mathrm{wp}$ vs $\Delta m^2_{21}$''.}
\label{fig:constrain_sigma_wp_up_dm}
\end{figure*}

The allowed region of (sin$^2 \theta_{12}$, $\sigma_\mathrm{wp}$) is shown in Fig. \ref{fig:constrain_sigma_wp_up_s212}. Our simulations show that the neutrino decoherence barely affect the measurement of $\theta_{12}$. As shown in Fig. \ref{fig:constrain_sigma_wp_up_s212}, larger $\sigma_\mathrm{wp}$ does not lead to larger value of sin$^2 \theta_{12}$, which ensures unbiased measurement on $\theta_{12}$. 
Besides, the future MBRO experiment(s) is also expected to provide precision measurements on the solar $\Delta m^2$ and atmospheric $\Delta m^2$. We also examine the potential neutrino wave-packet impact on the measurements of these two oscillation parameters. The results are shown in Fig. \ref{fig:constrain_sigma_wp_up_dm}. Similarly, these two panels ensure unbiased measurements on ($\Delta m^2_{31}$ + $\Delta m^2_{32}$) / 2 and $\Delta m^2_{21}$ under neutrino wave-packet treatment.

Our results suggest that although the future MBRO experiment(s) is expected to be more sensitive to the potential wave-packet impact, within the upper bounds of $\sigma_\mathrm{wp}$, the decoherence and dispersion effects are not large enough to cause significant damping signatures or modifications on the oscillation patterns. Therefore, we believe that the precision measurements on oscillation parameters at future MBRO experiment(s) are extremely safe even if neutrino is treated as wave packet. 

\section{The proposed extra detector}\label{sec5}

In section \ref{sec3}, we discussed how the statistical, systematic uncertainties and the resolution affected the sensitivities of MBRO in constraining $\sigma_\mathrm{wp}$. As a result, the larger statistics, the smaller systematic uncertainty and the nicer energy resolution can always provide a better sensitivity. On the other hand, building an extra detector could also be a potential approch to improving the sensitivity. In reality, the JUNO experiment \cite{An:2015jdp} is planning to build a near detector 30-35 m from a European pressure water reactor (EPR) of thermal power 4.6 GW$_\text{th}$, JUNO-TAO \cite{JUNO-TAO2}. In principle, such a small near detector is not sensitive to the decoherence effect since the damping factor in Eq. (\ref{Prob_inf_1st} depends on the baseline $L$). 

However, if an extra detector is built with a baseline of > 10 km, it maybe act on improving the sensitivity on constraining $\sigma_\mathrm{wp}$. Recently, there have been studies \cite{Wang:2016vua, Ciuffoli:2012bp, 2AD} suggest that bulding an additional detector at the intermediate baseline (around 10 to 40 km) can provide extra sensitivities for the neutrino MH resolution in the future MBRO experiment, since such detector is expected to be able to reduce the correlated uncertainties. Based on the proposal of an extra detector, we further investigate its potential benefits in studying the neutrino wave-packet impact.

\subsection{Comparing the sensitivities with single and multiple detectors}\label{sec5.1}

We use a similar setup of the extra detector as suggested by Reference \cite{Wang:2016vua}: a 4 kton detector with 3\% energy resolution, located at baseline of 12.5 km. However, in our simulation, we assume that the extra detector is not identical to the original one and thus will cause uncorrelated uncertainties, which is different with the assumption in Reference \cite{Wang:2016vua} and also other literatures \cite{Ciuffoli:2012bp, 2AD}. We believe that building an identical far and near detector is not feasible because any far detector capable of determining the mass hierarchy is quite large with a unique geometry, and there will therefore be many uncorrelated uncertainties to deal with. Additionally, if a near detector starts data taking after the far detector, this could introduce additional uncorrelated uncertainties. We believe that the proposed extra detector is not used to cancel the correlated uncertainties such as shape uncertainties. Moreover, as shown in Fig. \ref{fig:constrain_sigma_wp_ShapeEr_FD_only}, it is unlikely that the improvements on systematic uncertainties could significantly improve the sensitivity on constraining the parameter $\sigma_\mathrm{wp}$. In our simulations, we not only study the case of correlated shape uncertainties, but also examine the scenario that the shape uncertainties are uncorrelated and cannot be cancelled.

\begin{figure*}[!htbp]
\centering
 \includegraphics[scale=0.62]{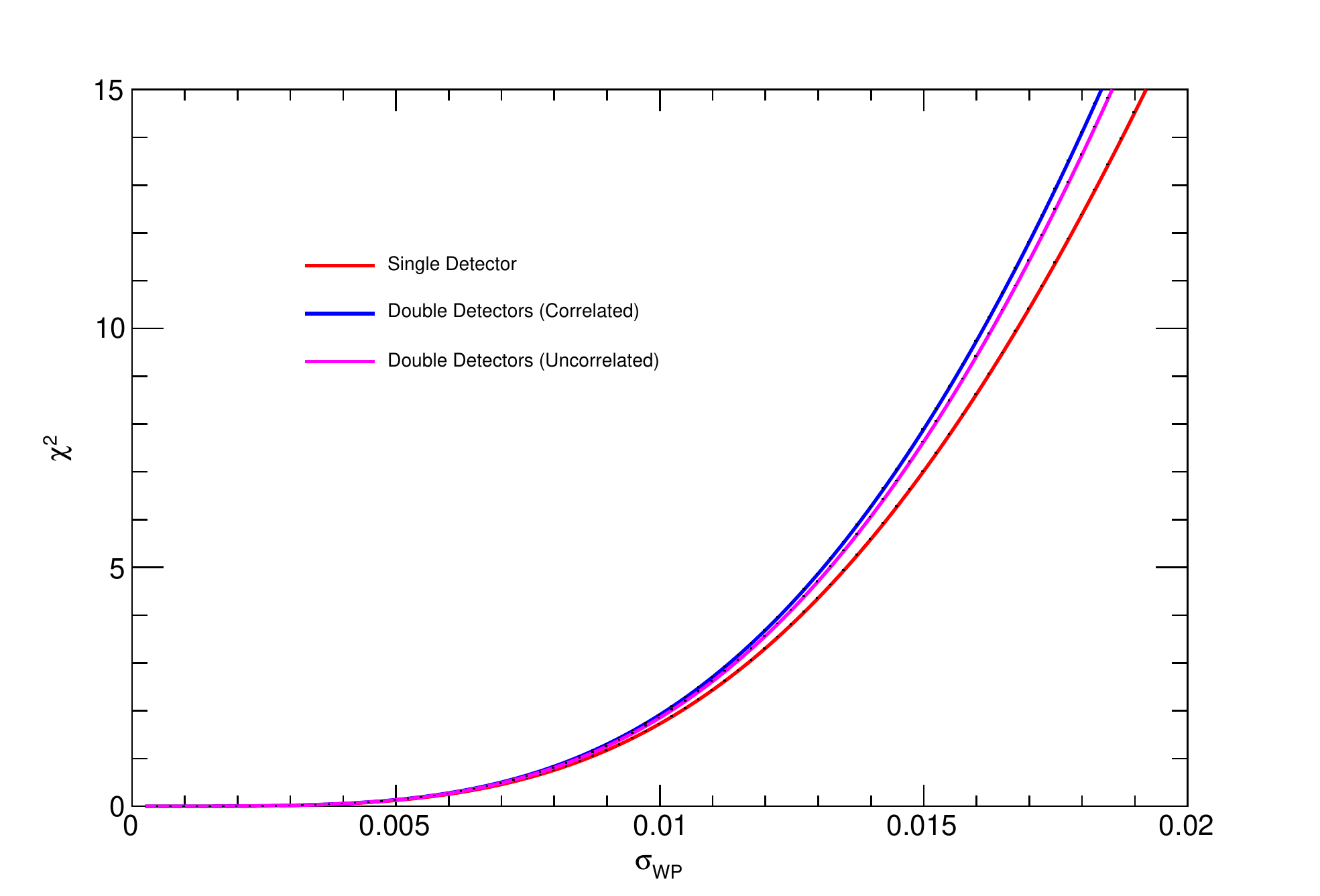} 
\caption{$\chi^2$ vs $\sigma_\mathrm{wp}$ for the single and double detector configurations. The red curve reveals the sensitivity of single detector at baseline of 52.5 km; The blue curve represents the case of two detectors, assuming the shape uncertainties are correlated and can be cancelled; The magenta curve corresponds to the assumption that the shape uncertainties are uncorrelated and cannot be canceled}
\label{fig:single_and_double_det}
\end{figure*}

We probe the potential benefits of an additional detector, as Reference \cite{Wang:2016vua} suggests that extra detector could give rise to extra sensitivity on the determination of neutrino mass ordering. It is because an intermediate baseline ($\sim$ 10 km) detector could provide extra information on the value of $\Delta m_{ee}^2$ in the analysis, which is important in the neutrino MH resolution. The sensitivity of constraining $\sigma_\mathrm{wp}$ with single and multiple detectors is shown in Fig. \ref{fig:single_and_double_det}. We compare three different scenarios: single medium-baseline detector (red curve), double detector with correlated (blue curve) and uncorrelated shape uncertainties (magenta curve).

Fig. \ref{fig:single_and_double_det} shows that the proposed extra detector could not significantly improve the sensitivity, no matter the shape uncertainties of two detectors are assumed to be correlated or uncorrelated. The blue curve and magenta curve are close to each other because the shape uncertainties are not crucial in the study of neutrino wave-packet impact, as illustrated in Fig. \ref{fig:constrain_sigma_wp_ShapeEr_FD_only}. The tiny difference between red curve and magenta (or blue) curve shows the extra sensitivity provided by the extra detector. Reference \cite{Wang:2016vua} suggests that the optimal baseline of extra detector should be around 12.5 km as in such location it could provide largest extra mass hierarchy sensitivity. However, the optimal location of the extra detector in studying decoherence effect could be different. 

\subsection{The optimal baseline of the extra detector}\label{sec5.2}
\begin{figure*}[!htbp]
\centering
 \includegraphics[scale=0.62]{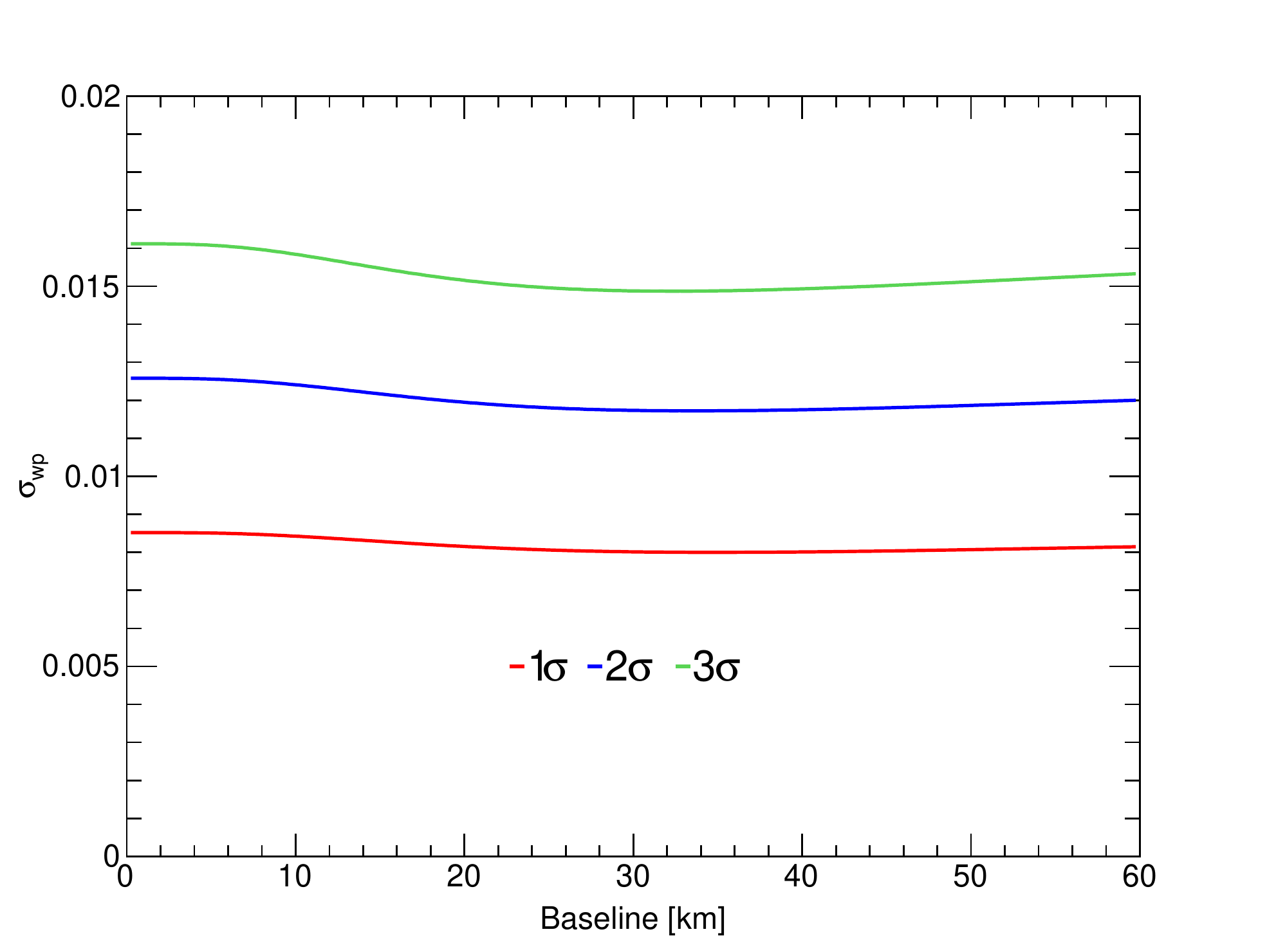} 
\caption{The 1, 2, 3 $\sigma$ C.L upper bound on $\sigma_\mathrm{wp}$ as a function of baseline of the extra detector.}
\label{fig:constrain_sigma_wp_baseline}
\end{figure*}

In principle, a longer baseline could lead to better sensitivity of $\sigma_\mathrm{wp}$, since the decoherence effect is expected to be more significant at longer distance. However, longer baseline also corresponds to larger statistical uncertainties. We examine the impact of the baseline on the upper bounds of $\sigma_\mathrm{wp}$, and search for the optimal location of the extra detector. The results of our numerical simulations are shown in Fig. \ref{fig:constrain_sigma_wp_baseline}.

According to our simulations, the upper bound on $\sigma_\mathrm{wp}$ slightly improves as the baseline of the extra detector increases. As mentioned before, longer traveling distance of the neutrinos is expected to enhance the decoherence effect. However, Fig. \ref{fig:constrain_sigma_wp_baseline} reveals that the impact of the extra detector is not large. It implies that the sensitivity of constraining $\sigma_\mathrm{wp}$ is mainly from the original medium-baseline detector, since the corresponding baseline is long and the target mass is large. Changing the location of the 4-ktons extra detector cannot improve the sensitivity of decoherence effect significantly.

\section{Conclusion}\label{sec6}
The neutrino wave-packet impact has been proved to be insignificant in the current reactor neutrino oscillation experiments. However, the plane-wave model of neutrino oscillation is only an approximation, and the wave-packet treatment is more general. Since the future MBRO experiment(s) could resolve multiple neutrino oscillations, it is expected to be an excellent platform to probe the potential decoherence effect or any other damping signatures in neutrino oscillations. 

In this article, a wave-packet treatment has been applied to study the $\bar\nu_e$ oscillations in the future MBRO experiment(s). The wave-packet treatment (with up to quadratic corrections) leads to decoherence, dispersion and delocalization effects, which modify the neutrino survival probability formula. In this article, numerical simulations for the future MBRO experiment(s) have been performed to probe the potential decoherence and dispersion effects.

Our simulations suggest that the 95\% C.L. allowed region of the parameter $\sigma_\mathrm{wp}$ is given by: 
\begin{equation}\label{95CL_allowed_region}
  7.51 \times 10^{-17} < \sigma_\mathrm{wp} < 0.0125,
\end{equation}
which is better than the current constraints from Daya Bay reactor neutrino experiment, especially the upper bound. 
Moreover, our simulations show that the wave-packet treatment does not lead to significant variations on the oscillation parameters ($\theta_{12}$,  $\Delta m^2_{21}$ and $\Delta m^2_{32}$). Within the 3 $\sigma$ C.L upper bound of $\sigma_\mathrm{wp}$, the wave-packet impact is not significant in the future MBRO experiment(s) and does not lead to significant shifts in the best-fit neutrino oscillation parameters.

We also discuss the experimental setups which affect the constraints on the parameter $\sigma_\mathrm{wp}$. Our simulations show that reducing the statistical uncertainties and shape uncertainties could improve the sensitivity, but the impact is not significant. Similarly, building an extra detector with intermediate baseline ($\sim$ 12 km) could also slightly improve the sensitivity, but the crucial factor is the detector energy resolution.

\section*{Acknowledgement}
The authors thank to Yasaman Farzan, Jiajun Liao, Dmitry Naumov and the JUNO collaboration Speakers Committee for informative discussions and suggestions. This study is supported in part by NSFC grant 11675273 and 2015M582453.

\bibliographystyle{elsarticle-num}
\bibliography{reference_reactor.bib} 

\end{document}